\documentclass[lettersize,journal]{IEEEtran}
\usepackage{amsmath,amsfonts}
\usepackage{algorithmic}
\usepackage{algorithm}
\usepackage{array}
\usepackage[caption=false,font=normalsize,labelfont=sf,textfont=sf]{subfig}
\usepackage{textcomp}
\usepackage{stfloats}
\usepackage{url}
\usepackage{verbatim}
\usepackage{graphicx}
\usepackage{cite}

\usepackage[pagebackref,breaklinks,colorlinks]{hyperref}
\usepackage{amssymb}
\usepackage{booktabs}
\usepackage{multirow}
\usepackage{pifont}

\begin{document}

\title{{CrossSpeech++: Cross-lingual Speech Synthesis with Decoupled Language and Speaker Generation}}

\author{Ji-Hoon Kim, Hong-Sun Yang, Yoon-Cheol Ju,  Il-Hwan Kim, Byeong-Yeol Kim, and Joon Son Chung
\thanks{Ji-Hoon Kim and Joon Son Chung are with School of Electric Engineering, Korea Advanced Institute of Science and Technology, Daejeon 34141, Republic of Korea (e-mail:jihoon@mm.kaist.ac.kr; joonson@kaist.ac.kr)

Hong-Sun Yang, Yoon-Cheol Ju, Il-Hwan Kim, and Byeong-Yeol Kim are with 42dot Inc., Seoul 06620, Republic of Korea (e-mail: hongsun.yang@42dot.ai; yooncheol.ju@42dot.ai; cutekih@gmail.com; byeongyeol.kim@42dot.ai)
}}

\markboth{Journal of \LaTeX\ Class Files,~Vol.~14, No.~8, August~2021}%
{Shell \MakeLowercase{\textit{et al.}}: A Sample Article Using IEEEtran.cls for IEEE Journals}


\maketitle
\begin{abstract} 
The goal of this work is to generate natural speech in multiple languages while maintaining the same speaker identity, a task known as cross-lingual speech synthesis. 
A key challenge of cross-lingual speech synthesis is the language-speaker entanglement problem, which causes the quality of cross-lingual systems to lag behind that of intra-lingual systems.
In this paper, we propose CrossSpeech++, which effectively disentangles language and speaker information and significantly improves the quality of cross-lingual speech synthesis. 
To this end, we break the complex speech generation pipeline into two simple components: language-dependent and speaker-dependent generators.
The language-dependent generator produces linguistic variations that are not biased by specific speaker attributes.
The speaker-dependent generator models acoustic variations that characterize speaker identity. 
By handling each type of information in separate modules, our method can effectively disentangle language and speaker representation.
We conduct extensive experiments using various metrics, and demonstrate that CrossSpeech++ achieves significant improvements in cross-lingual speech synthesis, outperforming existing methods by a large margin.
\end{abstract}


\section{Introduction}
\IEEEPARstart{I}{t} is believed that over 60 percent of the global population speaks at least two different languages~\cite{vince2016amazing,nam23_interspeech}.
In line with the recent trends in globalization, there has been growing interest in multi-lingual speech processing such as multi-lingual speech recognition~\cite{sahraeian2017crosslingual,choi2022distilling} or language identification~\cite{punjabi2021joint,park2023joint}.
In particular, cross-lingual Text-to-Speech (TTS) has attracted a large amount of attention due to its a range of applications, such as creating language educational content, developing conversational AI agents, and dubbing foreign movies.

Cross-lingual TTS focuses on generating natural-sounding speech in multiple languages while preserving the unique voice characteristics of the target speaker (e.g., synthesizing fluent Korean, Chinese and Japanese speech in the voice of Joe Biden). 
However, compared to intra-lingual TTS, which achieves almost human-like generation quality, the quality of cross-lingual TTS still lags far behind~\cite{lancucki2021fastpitch,du2023speaker,miao2024efficienttts}. 
One main challenge that degrades the generation quality of cross-lingual TTS is the language-speaker entanglement problem.
Specifically, since it is common for each speaker in a training dataset to speak only one language, there is a substantial risk of speaker identity becoming intertwined with language information during the training process.
In the extreme scenario where there is only a single speaker per language in the training data, the language identity perfectly matches the speaker identity.
These entangled representations hinder natural cross-lingual speech generation when the language identity is switched during the inference process, leading to unexpected speaker characteristics or unnatural pronunciation in the generated cross-lingual speech.

Numerous attempts have been made to disentangle language and speaker representations during training. 
Instead of using language-dependent text representation (e.g., graphemes), some works explore text representations which can be generalized cross multiple languages~\cite{li2019bytes,zhan21_interspeech,lux2022language}. 
Other works leverage domain generalization training techniques such as domain adversarial training~\cite{zhang2019learning} or mutual information minimization~\cite{xin2021disentangled} or information bottleneck methods~\cite{cong2023genertts}.
More recently, other works have utilized Self-Supervised Learning (SSL) speech representations based on the finding that SSL features capture only specific aspects of speech~\cite{liu23d_interspeech,gong2023zmm}.
Although previous studies have focused on decomposing language and speaker information, the decomposition is limited to the input features and does not fully address the entanglement problem. 
In other words, even if language and speaker representations are separated in the input token space, they are expected to be recombined when generating acoustic representations. 
This reintegration of separated representations prevents the synthesis of natural cross-lingual speech.

\begin{figure*}[ht]
    \centering
    \includegraphics[width=0.9\textwidth]{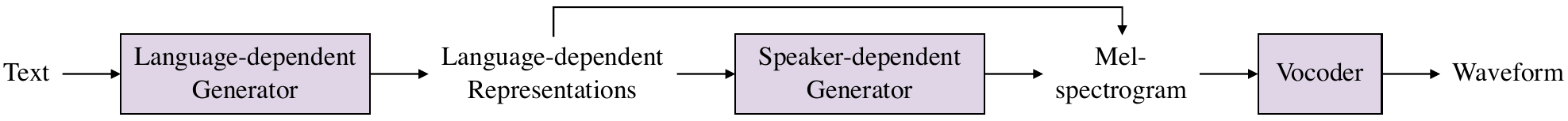}
    \caption{CrossSpeech++ operates as follows: From text inputs, the language-dependent generator produces language-dependent representations that capture
    linguistic characteristics drived solely from text inputs.
    The following speaker-dependent generator colorize speaker-specific attributes, and the mel-spectrogram is produced by summing representations from both generators. 
    The output mel-spectrogram is then converted to an audible waveform by a pre-trained neural vocoder.}
    \label{fig:concept}
    \vspace{-3.5mm}
\end{figure*}

In this paper, we propose CrossSpeech++ which improves the quality of synthesized cross-lingual speech by decomposing language and speaker information in the output acoustic feature space.
As depicted in Fig.~\ref{fig:concept}, CrossSpeech++ breaks intricate speech generation pipeline into two simple generator: the Language-dependent Generator (LDG) and the Speaker-dependent Generator (SDG), each of which produces the corresponding representations in the output feature space.
The language-dependent representations capture linguistic variation in speech, such as pronunciation and intonation, while speaker-dependent representations characterize speaker attributes such as timbre and pitch.

{Specifically, the LDG includes three components: Mix Dynamic Speaker Layer Normalization (MDSLN), the Language-Dependent Variance (LDV) adaptor, and the linguistic adaptor. 
MDSLN modulates text features with randomly mixed speaker information, mitigating language-speaker entanglement.
The LDV adaptor and linguistic adaptor model linguistic-related variations, enhancing robust cross-lingual speech generation.}
Similarly, the SDG comprises two modules: Dynamic Speaker Layer Normalization (DSLN) and the Speaker-Dependent Variance (SDV) adaptor.
Different from MDSLN, DSLN conveys speaker information, ensuring accurate speaker identity.
The SDV adaptor introduces speaker-specific acoustic variations, which are crucial for generating natural prosody.

We conduct extensive experiments to validate the effectiveness of CrossSpeech++, including subjective and objective evaluation metrics.
The results demonstrate that CrossSpeech++ significantly improves the quality of generated cross-lingual speech, in terms of both subjective and objective evaluation metrics.
The synthesized audio samples can be found on our demo page\footnote{\url{https://mm.kaist.ac.kr/projects/CrossSpeechpp}}.

\section{Related Works}
\subsection{Speech Synthesis}
Speech synthesis (text-to-speech, TTS), the process of synthesizing human speech from text, has a long history of innovation.
With the development of deep neural networks, recent deep-learning based TTS models have shown remarkable speech quality compared to early concatenative~\cite{hunt1996unit} and statistical methods~\cite{black2007statistical}, reaching speech quality close to that of real human utterance~\cite{tan2024naturalspeech}.
Typically, these methods involve converting a text sequence into intermediate acoustic representations and then transforming them to an audible waveform using either an external vocoder~\cite{matsubara2023harmonic,nguyen2024fregrad} or an internal decoder~\cite{ju2022trinitts,lee2022hierspeech}.
They employ various backbone networks such as dilated CNN~\cite{oord2016wavenet}, 
RNN~\cite{arik2017deep,shen2018natural}, and feed forward transformer~\cite{ren2021fastspeech,mehta2024matcha}.

Recent advancements have prompted TTS research to explore various topics such as multi-speaker~\cite{chen2020multispeech}, lightweight~\cite{luo2021lightspeech}, and cross-lingual TTS~\cite{zhang2019learning}.
Among these topics, cross-lingual TTS, in particular, demonstrates inferior synthetic quality compared to intra-lingual TTS mainly due to the language-speaker entanglement problem. 
In this paper, we focus on improving the quality of cross-lingual TTS to achieve high-quality speech synthesis on par with intra-lingual TTS.

\subsection{Cross-lingual Speech Synthesis}
Cross-lingual TTS, a branch of TTS, aims to produce natural speech in multiple languages while maintaining the same speaker identity.
In comparison to intra-lingual TTS, the quality of cross-lingual TTS remains weak due to the challenges in producing accurate speaker timbre and natural-sounding foreign accents.
The inferior quality of cross-lingual TTS primarily arises from the language-speaker entanglement issue~\cite{zhang2019learning}.
To address this, numerous efforts have been made, which generally fall into two broad categories: one is to leverage language-agnostic input representation, and the other seeks to learn disentangled representation.

Instead of relying on language-dependent input representations such as graphemes, some works present their cross-lingual systems based on language-independent input representations, which can be commonly used for multiple languages.
Zhan \textit{et al.}~\cite{zhan21_interspeech} employ the International Phonetic Alphabet (IPA) and demonstrate its superiority over language-dependent phonemes in enhancing the quality of cross-lingual TTS.
Li \textit{et al.}~\cite{li2019bytes} adopt UTF-8 byte representations for encoding typographic information, distancing their system from language-specific constraints.
Staib \textit{et al.}~\cite{staib2020phonological} and Lux \& Vu~\cite{lux2022language} utilize input representations derived from IPA articulation, specifically designed to maintain consistent topology across different languages.
Furthermore, Saeki \textit{et al.}~\cite{saeki2024text} explore the cross-lingual transferability based on BERT-like multilingual language model~\cite{kenton2019bert}, pushing the boundaries of cross-lingual transfer in TTS. 

Another approach presents training strategy to learn disentangled language and speaker representations.
Zhang \textit{et al.}~\cite{zhang2019learning} employ domain adversarial training~\cite{ganin2016domain} to prevent the leakage of speaker information from text encoding.
Xin \textit{et al.}~\cite{xin2021disentangled} leverage mutual information minimization loss~\cite{sanchez2020learning} to remove common attributes between language and speaker representation.
SANE-TTS~\cite{cho2022sane} proposes the speaker regularization loss to avoid speaker bias in text duration predictor, and GenerTTS~\cite{cong2023genertts} incorporates an information bottleneck to disentangle timbre and speaker style.
More recently, DSE-TTS~\cite{liu23d_interspeech} and ZMM-TTS~\cite{gong2023zmm} utilize SSL-based speech representations, as their discretized features contain less speaker-dependent information.
Although these previous works have attempted to address the language-speaker entanglement problem, the level of disentanglement has been limited to the input token space.
{
To address this, in our previous work, CrossSpeech~\cite{kim2023crossspeech}, we explicitly divide the speech generation pipeline into language-related and speaker-related components, with each generating the corresponding representation in the output feature space. 
In this paper, we further explore the advantages of splitting the speech generation to develop a more natural cross-lingual TTS system. Moreover, we incorporate additional language- and speaker-specific attributes to further enhance generation quality.}

\subsection{Domain Generalization}
Domain generalization focuses on training models to perform well on any unseen domain, which is not accessible during training. 
Numerous seminal works have collectively advanced the field of domain generalization. 
Many cross-lingual TTS methods leverage domain generalization techniques with the aim of enabling the models to effectively generalize well to unseen language-speaker combinations.

One of the foundational works in domain generalization is domain adversarial training (DAT)~\cite{ganin2016domain}.
DAT introduces a gradient reversal layer that ensures the feature extractor learns to produce features indistinguisable across multiple domains by reversing the gradients from a domain discriminator during backpropagation.
Arjovsky \textit{et al.}~\cite{arjovsky2019invariant} presents invariant risk minimization, a framework aims at learning domain-invariant predictors by leveraging the principle of risk invariance.
Zhou \textit{et al.}~\cite{zhou2021domain} propose a simple yet effective approach to domain generalization called MixStyle. 
This technique involves mixing feature statistics of training samples from different domains to generate new feature statistics that do not exist in the training data.
By doing so, MixStyle simulates domain shifts at the feature level, enabling the model to learn more generalized representations.
Motivated by this, in this work, we introduce a speaker-generalization module to prevent speaker bias in text embedding, mitigating the language-speaker entanglement problem in cross-lingual TTS.

\begin{figure*}[ht]
    \centering
    \includegraphics[width=0.85\textwidth]{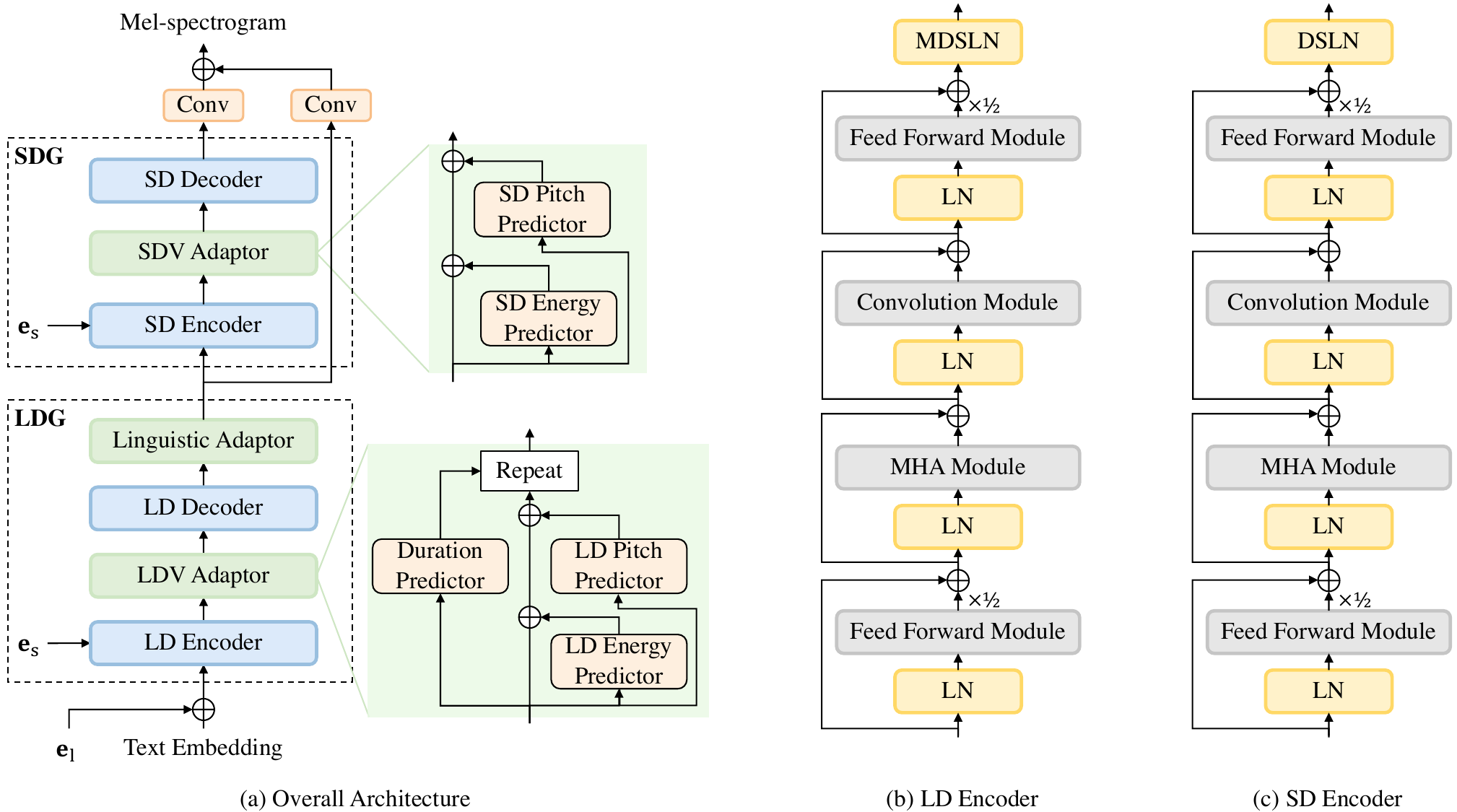}
    \caption{The overall architecutre of CrossSpeech++. ${\bf e}_\text{l}$ and ${\bf e}_\text{s}$ denote the language and speaker embeddings which are derived from trainable lookup tables. Detailed architectures of Language-dependent (LD) conformer encoder and Speaker-dependent (SD) conformer encoder are depicted in (b) and (c), respectively. 
    MHA means multi-head attention.
    We replace the final Layer Normalization (LN) in the conformer with Mix-dynamic Speaker Layer Normalization (MDSLN) in the LD encoder and Dynamic Speaker Layer Normalization (DSLN) in the SD encoder.
    }
    \label{fig:model}
    \vspace{-3.5mm}
\end{figure*}

\section{Model Architecture}
CrossSpeech++ is built upon FastPitch~\cite{lancucki2021fastpitch}, a non-autoregressive TTS model whose encoder and decoder are based on multiple feed-forward transformer blocks. 
It takes a text sequence ${\bf x}\in\mathbb{R}^{L}$ as input and produces a mel-spectrogram ${\bf y}\in\mathbb{R}^{T\times 80}$, where $L$ and $T$ denote the lengths of the text sequence and the output mel-spectrogram, respectively.
We adopt the online duration aligner~\cite{badlani2022one}, which allows ground truth durations to be obtained without external sources. 
This online aligner not only enables efficient training but also, more importantly, removes the dependency on pre-computed aligners for each language, which is highly beneficial for extending languages in cross-lingual TTS~\cite{badlani2022one}.
In addition, we replace the transformer with conformer blocks~\cite{gulati2020conformer} due to their capability to model rich features in a parameter-efficient way. 
To support multi-lingual and multi-speaker settings, we adopt trainable lookup tables for language and speaker.

An intuitive way to avoid language-speaker entanglement in cross-lingual TTS is to divide the generation pipeline into language-dependent and speaker-dependent parts~\cite{li2019feature,huang2022generspeech}.
As illustrated in Fig.~\ref{fig:model}, CrossSpeech++ breaks the speech generation pipeline into LDG and SDG, which model the language-dependent and speaker-dependent representations, respectively. 
Each generator includes multiple conformer blocks and other key components to obtain disentangled representations, which are described in the following sections.

\section{Language-dependent Generator}

In order to produce language-dependent representations, we specifically design the LDG to include MDSLN.
Additionally, to learn prosodic variations that are dependent only on linguistic information (e.g., pronunciation and intonation), we introduce the LDV adaptor and the linguistic adaptor. 
These collectively contribute to improving the quality of synthesized cross-lingual speech.

\subsection{Speaker Generalization}
\label{sec:spkgen}
The key to high-quality cross-lingual speech synthesis is to produce text features that are not biased toward any specific speaker.
To achieve this, we propose MDSLN, which is an extended module of DSLN~\cite{lee2022pvae}. 
DSLN adaptively modulates hidden features based on speaker statistics, rather than simply conditioning the speaker embeddings through summation or concatenation. 
Given hidden representations $\bf{h}$ and speaker embeddings $\bf{e}_\text{s}$, the speaker-conditioned representations are derived as follows:
\begin{equation}
    \text{DSLN}(\bf{h},{\bf e}_\text{s}) = \mathbf{W}({\bf e}_\text{s}) * \text{LN}(\bf{h}) + \mathbf{b}({\bf e}_\text{s}),
\end{equation}
where $*$ denotes 1D convolution, and LN refers to layer normalization. The normalized hidden feature space is then shifted according to speaker embedding statistics, the filter weight $\mathbf{W}$ and bias $\mathbf{b}$, which are predicted by a single linear layer using $\bf{e}_\text{s}$ as input.

\begin{figure}[t]
    \centering
    \includegraphics[width=0.8\columnwidth]{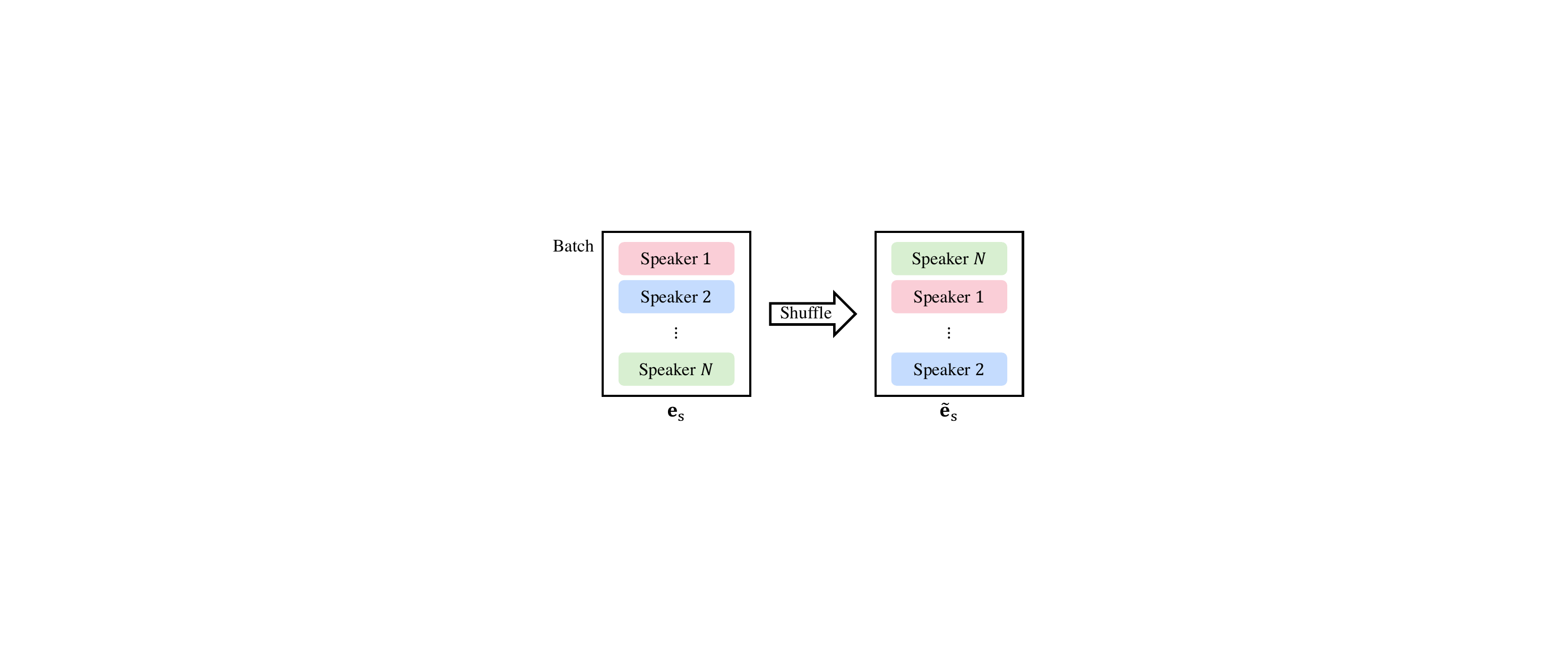}
    \caption{Batch-wise shuffle operation. ${\bf e}_\text{s}$ is the speaker embeddings and $\Tilde{{\bf e}}_\text{s}$ denotes the shuffled speaker embeddings.}
    \label{fig:shuffle}
    \vspace{-3.5mm}
\end{figure}

Intuitively, the model can learn speaker-generalizable text features when the text features are continuously adapted by a random speaker during training.
This allows the model to selectively capture essential text-related attributes, apart from speaker-related information.
The adaptation is achieved by conditioning the text representation with random speaker information. 
Inspired by recent works~\cite{zhou2021domain, jin2021style, huang2022generspeech}, we introduce MDSLN to continuously refine the text sequence with random speaker information by mixing speaker distributions in the training data, which can be formulated as follows:
\begin{equation}
    \text{MDSLN}(\bf{h},{\bf e}_\text{s}) = \mathbf{W}_\text{mix}({\bf e}_\text{s}) * \text{LN}(\bf{h}) + \mathbf{b}_\text{mix}({\bf e}_\text{s}),
\end{equation}
where $\bf{W}_\text{mix}$ and $\bf{b}_\text{mix}$ represent filter weight and bias for a randomly mixed speaker distribution. 
The mixed speaker statistics can be calculated as follows:
\begin{equation}
    \mathbf{W}_{\text{mix}}({\bf e}_\text{s}) = \gamma \mathbf{W}({\bf e}_\text{s}) + (1-\gamma)\mathbf{W}(\tilde{{\bf e}}_\text{s}),
\end{equation}
\begin{equation}
    \mathbf{b}_{\text{mix}}({\bf e}_\text{s}) = \gamma \mathbf{b}({\bf e}_\text{s}) + (1-\gamma)\mathbf{b}(\tilde{{\bf e}}_\text{s}),
\end{equation}
where $\tilde{{\bf e}}_\text{s}$ is acquired by randomly shuffling ${\bf e}_\text{s}$ along the batch dimension (see Fig.~\ref{fig:shuffle}), and $\gamma$ is sampled from a Beta distribution: $\gamma \sim \text{Beta}(\alpha,\alpha)$ (we set $\alpha=2$ in our experiments). 
We substitute the LN at the end of the Language-dependent (LD) conformer encoder block with MDSLN.

\subsection{Language-dependent Variance Adaptor}
\label{sec:ldv}

Although it is crucial to model rich speech variations to synthesize expressive speech, predicting these variations in a cross-lingual scenario is challenging due to the combinations of languages and speakers that are unseen during training~\cite{ren2022prosospeech,oh2024diffprosody}. 
To address this issue, we introduce the LDV adaptor, which models text-driven speech variations, a common attribute across multiple speakers. 
This adaptor predicts binary pitch and energy variations, indicating whether these values rise or fall~\cite{zhan21_interspeech}.
The LDV adaptor consists of three components: a duration predictor, an LD pitch predictor, and an LD energy predictor, all sharing the same architecture. 
Pitch and energy values are embedded using a single 1D convolutional layer and are then added to the speaker-generalized text features. 
During training, we use the target values, while during inference, we rely on the predicted values.
The target duration value is obtained through an internal aligner~\cite{badlani2022one}, and targets for the LD pitch and energy predictors will be detailed in the following paragraphs.

\subsubsection{Pitch}

To obtain the target value for the LD pitch predictor, we first extract the ground truth pitch value for every frame using the {\tt pYIN} algorithm~\cite{mauch2014pyin}.
Since pitch is inherently speaker-dependent, we refer to the ground truth pitch sequence as the speaker-dependent pitch sequence, denoted as ${\bf {p}^{(s)}} \in \mathbb{R}^{T}$.
We average $\bf p^{(s)}$ across each input text token using the ground truth duration, and convert the averaged sequence (denoted as ${\bf \bar{ p}^{(s)}} \in \mathbb{R}^{L}$) into a binary sequence. 
This results in the language-dependent (speaker-independent) target pitch sequence ${\bf p^{(l)}} \in \mathbb{R}^{L}$. 
The conversion to a binary sequence is defined as follows:
\begin{equation}
    {p}^{(l)}_i = 
    \begin{cases}
        1, & \mbox{if}~~{\bar{p}}^{(s)}_{i-1}<\bar{p}^{(s)}_i,\\
        0, & \mbox{otherwise},
    \end{cases}
\end{equation}
where $\bar{p}^{(s)}_i$ denotes the $i^{th}$ value of $\bf \bar{p}^{(s)}$, and $p^{(l)}_i$ represents the $i^{th}$ value of ${\bf p^{(l)}}$ for $i \in \{1, 2, 3, \ldots, L\}$.
Using $\bf{p}^{(l)}$ as the target, the LD pitch predictor is optimized with a binary cross-entropy loss:
\begin{equation}
    \mathcal{L}_\text{LDP} = - \sum^{L}_{i=1}\big[p^{(l)}_i \text{log}\hat{p}^{(l)}_i + (1-p^{(l)}_i) \text{log}(1-\hat{p}^{(l)}_i)\big],
\end{equation}
where $\hat{p}_{i}^{(l)}$ denotes the $i^{th}$ predicted language-dependent pitch. 

\subsubsection{Energy} 

We extract the speaker-dependent energy, ${\bf e^{(s)}}$, by taking an average from a target mel-spectrogram along the frequency axis~\cite{choi2021neural}. 
Similar to pitch, we average ${\bf {e}^{(s)}}\in\mathbb{R}^{T}$ over every text token and compute the language-dependent energy ${\bf e^{(l)}}\in\mathbb{R}^{L}$ by transforming the averaged sequence into a binary sequence.
The Language-dependent Energy (LDE) predictor is also trained through binary cross-entropy loss between the predicted and target LD energy sequence:
\begin{equation}
    \mathcal{L}_\text{LDE} = - \sum^{L}_{i=1}\big[e^{(l)}_i \text{log}\hat{e}^{(l)}_i + (1-e^{(l)}_i) \text{log}(1-\hat{e}^{(l)}_i)\big],
\end{equation}
where $e_i^{(l)}$ and $\hat{e}_{i}^{(l)}$ represent the $i^{th}$ the target and the predicted language-dependent energy value, respectively.
 
The enriched hidden sequence is upsampled according to the token durations and then fed to the Language-dependent (LD) conformer decoder. 
Note that the duration predictor in CrossSpeech++ learns \textit{general} duration information because it takes speaker-generalized representation as an input.
As proven in the recent study~\cite{cho2022sane}, this leads to predicting token duration independently from speaker identity and stabilizes the duration prediction in cross-lingual TTS.

\subsection{Linguistic Adaptor}
Text-dependent speech variations likely encompass a variety of complex characteristics, motivating us to construct a linguistic adaptor that further enriches text-related acoustic attributes beyond LD pitch and energy.
The linguistic adaptor shares the same teacher-forcing strategy as the LDV adaptor but it has a distinct target features deliberately designed to contain elaborate linguistic features independent of speaker-specific characteristics.
The linguistic adaptor contains linguistic predictor that directly estimate the target linguistic features {from the output of the LD decoder}, and it is trained with L1 loss:
\begin{equation}
    \mathcal{L}_\text{L} = ||{\bf{l}}-{\bf{\hat{l}}} ||_1,
\end{equation}
where ${\bf{l}}$ and ${\bf{\hat{l}}}$ refer to the target and predicted linguistic features, respectively.

\begin{figure}[t]
    \centering
    \includegraphics[width=0.5\columnwidth]{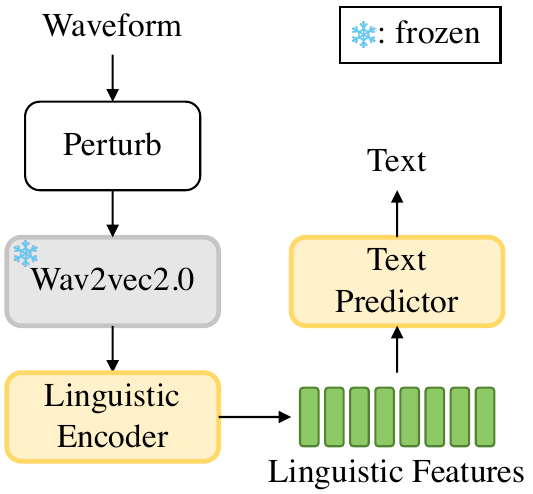}
    \caption{A pipeline for extracting the target linguistic features from waveform. 
    To eliminate speaker-related features, the perturbed waveform is input to a multilingual wav2vec2.0. 
    The following linguistic encoder extracts solely linguistic features, which are fed into a text predictor. 
    }
    \label{fig:linguistic}
    \vspace{-3.5mm}
\end{figure}

As illustrated in Fig.~\ref{fig:linguistic}, we extract the target linguistic features by leveraging self-supervised speech representations. 
Previous studies have shown that representations from the SSL speech models contain comprehensive information, with each layer exhibiting different aspects of speech~\cite{fan2020exploring,kim2024let}.
Based on empirical observations, we decide to utilize the last hidden feature of MMS~\cite{pratap2024scaling}, a wav2vec2.0 model~\cite{baevski2020wav2vec} pre-trained on over 500k hours of speech across 1,400 languages.
We also employ information perturbation techniques~\cite{choi2021neural} that can remove speaker-dependent information in the waveform, such as formants, pitch, and frequency response, through a series of formant shifting, pitch randomization, and random frequency shaping functions.
Subsequently, the linguistic encoder extracts the target linguistic features, which are fed into an auxiliary text predictor to strengthen linguistic characteristics~\cite{lee2022hierspeech}.
Both the linguistic encoder and the text predictor are optimized with connectionist temporal classification (CTC) loss~\cite{graves2006connectionist} between the text sequence ${\bf x}$ and the linguistic features ${\bf z}$: $\mathcal{L}_\text{CTC} = -\text{log}P({\bf x}|{\bf z})$~\cite{lee2022hierspeech}.
Note that the parameters of the pre-trained MMS are not updated.

\section{Speaker-dependent Generator}
To colorize speaker-specific attributes constituting one half of natural human speech, we construct an SDG that includes an SD encoder, an SDV adaptor, and an SD decoder.
SD encoder effectively aligns the language-dependent representations to the speaker identity with the help of DSLN~\cite{lee2022pvae}.
We stack conformer blocks for the SD encoder and replace the LN at the end of each conformer block with DSLN~\cite{lee2022pvae}.
The following SDV adapter consists of the speaker-dependent pitch (SDP) and energy (SDE) predictor. 
This adds speaker-specific speech variations such as formants and stress patterns.
We extract the speaker-dependent pitch ${\bf p^{(s)}}$ and energy ${\bf e^{(s)}}$ sequence 
as described in Sec.~\ref{sec:ldv}, and optimize the SD predictors using L1 loss:
\begin{equation}
    \mathcal{L}_\text{SDP} = ||{\bf{p}^{(s)}}-{\bf{\hat{p}}^{(s)}} ||_1,
\end{equation}
\begin{equation}
    \mathcal{L}_\text{SDE} = ||{\bf{e}^{(s)}}-{\bf{\hat{e}}^{(s)}} ||_1,
\end{equation}
where ${\bf{\hat{p}}^{(s)}}$ and ${\bf{\hat{e}}^{(s)}}$ denotes the predicted speaker-dependent pitch and energy sequences, respectively.
The speaker-dependent sequences are fed to the 1D convolutional layer and summed to the speaker-specific hidden feature.
The SD decoder then produces speaker-dependent acoustic representation.
The output mel spectrogram is generated by adding language-dependent and {speaker}-dependent features after they are projected through a single convolutional layer.

To sum up, the overall training objectives ($\mathcal{L}_\text{all}$) are given:
\begin{equation}
    \begin{split}
    \mathcal{L}_\text{all} = &\mathcal{L}_\text{mel} + \mathcal{L}_\text{align} + \lambda_\text{dur}\mathcal{L}_\text{dur} \\
    & + \lambda_\text{LDP}\mathcal{L}_\text{LDP} + \lambda_\text{LDE}\mathcal{L}_\text{LDE} + \lambda_\text{L}\mathcal{L}_\text{L} \\
    & + \lambda_\text{CTC}\mathcal{L}_\text{CTC} + \lambda_\text{SDP}\mathcal{L}_\text{SDP} + \lambda_\text{SDE}\mathcal{L}_\text{SDE}, \\
    \end{split}
\end{equation}
where $\mathcal{L}_\text{mel}$ means L1 loss between the target and the predicted mel-spectrogram, $\mathcal{L}_\text{align}$ denotes the alignment loss for the online aligner as described in~\cite{badlani2022one}. 
$\mathcal{L}_{dur}$ is L1 loss between the target and the predicted duration.
In our experiments, we fix $\lambda_\text{dur}=\lambda_\text{LDP}=\lambda_\text{LDE}=\lambda_\text{L}=\lambda_\text{CTC}=\lambda_\text{SDP}=\lambda_\text{SDE}=0.1$. 

\section{Experimental Settings}
\begin{table}[t]
  \centering
  \caption{Dataset description.}
  \begin{tabular}{cccc}
    \toprule
     Languages    & Source   &\#speakers &Hours  \\
    \cmidrule(lr){0-3}
    \multicolumn{1}{c}{\multirow{2}{*}{en-US}}  & LJSpeech~\cite{ljspeech17}  &1 &12.228    \\
    \multicolumn{1}{c}{} & VCTK~\cite{vctk19} &3 & ~0.581\\ \cmidrule(lr){0-3}
    \multicolumn{1}{c}{\multirow{2}{*}{zh-CN}} & Databaker~\cite{BIAOBEI} &1 &10.080 \\ 
    \multicolumn{1}{c}{} &AIShell3~\cite{shi21c_interspeech} &9 & ~3.614 \\ \cmidrule(lr){0-3}
    \multicolumn{1}{c}{\multirow{2}{*}{ja-JP}} &CSS10~\cite{park19c_interspeech} &1 &~6.563\\
    \multicolumn{1}{c}{} &JSUT~\cite{sonobe2017jsut} &1 &~7.458 \\ \cmidrule(lr){0-3}
    ko-KR &koMulti~\cite{komulti} &6 &14.095 \\
    \bottomrule
  \end{tabular}
  \vspace{-3.5mm}
  \label{tab:data}
\end{table}

\subsection{Dataset}
We conduct experiments on the mixture of the monolingual dataset in four languages: English (en-US), Chinese (zh-CN), Japanese (ja-JP), and Korean (ko-KR) as detailed in Table~\ref{tab:data}. 
Since all the datasets have different environments, we resample all the audio to 16kHz and convert the corresponding transcripts to IPA symbols~\cite{bernard2021phonemizer}.
In our experiments, the dataset is split into 80\%-10\%-10\% for training, validation, and test sets across all speakers.
80 bins mel-spectrogram is transformed with a window size of 1280, a hop size of 320, and
Fourier transform size of 1280.

\subsection{Model Configuration}
All the encoders and decoders in our method are based on conformer blocks. 
Except for the LD encoder, which consists of 4 conformer blocks, the other modules are composed of 2 conformer blocks each.
Each conformer block is designed with a hidden dimension of 192 and a single attention head. 
We also set a hidden dimension of 192 for the language and speaker lookup tables, where each language and speaker ID is converted into a 192-dimensional embedding vector. 
The variance predictors share the same architecture, which consists of two 1D convolutional layers with ReLU activation, each followed by layer normalization and a dropout layer, as in FastSpeech2~\cite{ren2021fastspeech}. 
Following recent work on voice conversion~\cite{choi2022nansy++}, the linguistic encoder includes a Convolutional Gated Linear Unit (ConvGLU)~\cite{dauphin2017language}, and we add layer normalization at the end of the linguistic encoder to stabilize the linguistic feature prediction pipeline. 
We utilize pre-trained MMS from {\tt Hugging Face}\footnote{https://huggingface.co/facebook/mms-300m}, and our text predictor consists of 2 conformer blocks followed by a single projection layer.
{The total number of learnable parameters is 12M.}

\subsection{Training Details}
CrossSpeech++ is trained for 500 epochs on 8 NVIDIA A6000 GPUs with a batch size of 128. We use the AdamW optimizer with $\beta_1=0.8$, $\beta_2=0.99$, $\epsilon=10^{-9}$, and an initial learning rate of $2 \times 10^{-4}$, decayed by $0.999875$ per epoch. 
Gradients are accumulated and the optimizer steps after every two batches to enhance training efficiency.

\subsection{Baseline Methods}
CrossSpeech++ is compared against recent cross-lingual TTS systems. 
All the systems are trained and evaluated with the same configurations, including training and test datasets. The output mel-spectrogram is converted to an audible waveform by pre-trained Fre-GAN~\cite{kim21f_interspeech} vocoder.
\begin{itemize}
    \item \textbf{FastPitch (FP)}~\cite{lancucki2021fastpitch} is the backbone network of CrossSpeech++. 
    We follow the official implementation of FastPitch\footnote{https://github.com/NVIDIA/DeepLearningExamples/tree/master/PyTorch/\\SpeechSynthesis/FastPitch} with slight modifications.
    We incorporate trainable lookup tables to support multiple speakers and languages, and adopt the online duration aligner~\cite{badlani2022one}. 
    \item \textbf{FP + DAT}~\cite{zhang2019learning} adopts domain adversarial training (DAT) based on FastPitch.
    Given that the DAT speaker classifier proposed by Zhou \textit{et al.}~\cite{zhou2021domain} can be easily applied to other systems, we integrated this DAT classifier into FastPitch.
    \item \textbf{FP + DAT} + $\mathbb{\mathcal{L}}_\text{\bf reg}$~\cite{cho2022sane} leverages speaker regularization loss ($\mathcal{L}_\text{reg}$) along with the DAT classifier as in SANE-TTS~\cite{cho2022sane}. 
    Since the speaker regularization loss stabilizes the duration prediction process in non-autoregressive TTS systems, it can be applied to any non-autoregressive system that adopts a duration predictor. 
    Therefore, we choose FastPitch as the backbone network.
    \item \textbf{CrossSpeech}~\cite{kim2023crossspeech} is our previous work and serves as a strong and comparable baseline to our current model.
    While it shares similarities with CrossSpeech++ in dividing the speech generation process into speaker-independent and speaker-dependent modules, CrossSpeech++ introduces more speech variation (i.e., LD and SD energy). More importantly, it incorporates SSL-based linguistic information which is the key to improving speech quality.
\end{itemize}

\begin{table*}[t]
\centering
\caption{Evaluation results. MOS and SMOS are presented with $95\%$ confidence interval. UTMOS is an automatic prediction model for MOS. SECS denotes speaker embedding cosine similarity and CER refers to character error rate. Lower is better for CER, and higher is better for the other metrics. The bold value represents the best score for each metric.}
\resizebox{0.99\textwidth}{!}
{
\begin{tabular}{lcccccccccc}
\toprule
\multirow{2}{*}{\vspace{-1mm}\bfseries Method}     & \multicolumn{5}{c}{\bfseries Cross-lingual} & \multicolumn{5}{c}{\bfseries Intra-lingual} \\ \cmidrule(lr){2-6}\cmidrule(lr){7-11}
      & MOS$\uparrow$   &SMOS$\uparrow$ &UTMOS$\uparrow$  &SECS$\uparrow$ &CER$\downarrow$ & MOS$\uparrow$ &SMOS$\uparrow$ &UTMOS$\uparrow$  &SECS$\uparrow$ &CER$\downarrow$   \\ \cmidrule(lr){1-11}
Ground Truth    &$-$       &$-$        &$-$ &$-$ &$-$ &$4.55\pm0.06$   & $4.93\pm0.05$ & $4.052$  &$0.793$ &$8.57$       \\
Vocoded &$-$       &$-$     &$-$      &$-$ &$-$ &$4.38\pm0.09$     & $4.78\pm0.06$   & $3.833$ & $0.791$ &$8.88$      \\\cmidrule(lr){1-11}
FP~\cite{lancucki2021fastpitch} &$3.88\pm0.07$ &$3.23\pm0.09$  &$3.474$  &$0.711$ &$14.47$  & $3.65\pm0.09$     &$3.92\pm0.10$ &$3.074$ & $0.773$ &$11.11$    \\\cmidrule(lr){1-11}
FP+DAT~\cite{zhang2019learning} &$3.55\pm0.09$ &$3.66\pm0.09$ &$3.468$ &$0.738$ &$14.84$ & $3.49\pm0.10$ & $3.88\pm0.11$ &$3.086$ &$0.776$ &$10.73$ \\
FP+DAT+$\mathcal{L}_{reg}$~\cite{cho2022sane} &$3.71\pm0.11$ &$3.65\pm0.08$  &$3.490$ & $0.756$ &$14.26$ & $3.43\pm{0.11}$ & $3.82\pm0.09$ &$3.087$ &$0.776$ &$10.65$ \\ 
CrossSpeech~\cite{kim2023crossspeech}  &$3.93\pm0.08$ &$\bf 3.87\pm0.07$ &$3.279$  &$\bf 0.776$ &$16.15$ &$3.56\pm0.12$ & ${3.86}\pm0.09$ &$3.039$ & $\bf 0.781$ & $11.26$ \\ \cmidrule(lr){1-11}
\textbf{CrossSpeech++} &$\bf 4.06\pm0.09$ &$3.82\pm0.10$ &$\bf 3.791$  &${0.761}$ &$\bf 13.35$ & $\bf 3.85\pm0.09$ & $\bf 3.94\pm0.08$ &$\bf 3.343$ & $0.777$ &$\bf 10.14$\\
\bottomrule
\label{tab:compare}
\end{tabular}
}
\vspace{-5mm}
\end{table*}

\subsection{Evaluation Metrics}
We assessed the effectiveness of our method using extensive evaluation metrics, including both subjective and objective measures. 
We used 50 random speech clips for subjective evaluation (i.e., MOS and SMOS), and 300 samples for objective evaluations (UTMOS, SECS, and CER).
\begin{itemize}
    \item \textbf{MOS} stands for Mean Opinion Score. 
    To evaluate the naturalness of audio, we performed a MOS test in which 30 domain-expert subjects were asked to rate the naturalness on a scale from 1 to 5. 
    Speech naturalness includes audio clarity and pronunciation accuracy.
    \item \textbf{SMOS} denotes Similarity Mean Opinion Score. 
    Similar to MOS, 30 raters assess the speaker similarity of speech pairs. 
    The raters were instructed to focus solely on the voice similarity to the target speaker; high scores are given if the voices are similar, even if the quality of speech is degraded.
    \item \textbf{UTMOS}\cite{saeki2022utmos} is an automatic MOS prediction neural network. 
    While subjective evaluation is regarded as the gold standard in assessing speech naturalness\cite{black2005blizzard}, it requires high costs in terms of both time and money. 
    As a remedy to this, UTMOS has been widely used because of its effectiveness in estimating subjective scores~\cite{wang2024usat,sun2024dual,kameoka2024voicegrad}.
    \item \textbf{SECS} denotes Speaker Embedding Cosine Similarity. 
    It measures how similar the speaker characteristics of the generated speech are to those of the target speech.
    We extracted speaker representation using \texttt{Resemblyzer}\footnote{https://github.com/resemble-ai/Resemblyzer} from generated and the actual speech, then computed the cosine simliarity between them.
    \item \textbf{CER} stands for Character Error Rate, which measures the intelligibility of speech by comparing the predicted text of speech to the target text sequence.
    We obtained the transcriptions of speech using a publicly available automatic speech recognition (ASR) system~\cite{radford2023robust} that is pre-trained on 680k hours of speech from 99 languages.
\end{itemize}

\section{Results and Analysis}
\subsection{Quality Comparison}
To show the effectiveness of CrossSpeech++, we compare the generation performance of CrossSpeech++ against that of recent cross-lingual TTS models on both cross-lingual and intra-lingual scenarios.
{For cross-lingual evaluation, we randomly sample four representative speakers per language, while all speaker IDs are used for intra-lingual evaluation.}
The results are listed in Table~\ref{tab:compare}.
Above all, CrossSpeech++ achieves significant improvements in cross-lingual speech.
In cross-lingual TTS, CrossSpeech++ obtains the best scroes in MOS as well as UTMOS and CER, which underscores the ability of CrossSpeech++ to generate highly natural speech.
While CrossSpeech++ shows a slight decrease in similarity scores (SMOS and SECS) compared to our previous work, CrossSpeech, we posit that this difference is attributable to our method generating more precise pronunciation and accent driven by text inputs, which leads to a perceptible shift in speaker similarity to the target speaker using a different language.
CrossSpeech++ also demonstrates superior quality compared to the baselines in intra-lingual cases, confirming that our method is beneficial not only in cross-lingual settings but also in intra-lingual scenarios.

\subsection{Analysis on Linguistic Features}
To determine the optimal extraction pipeline for the target linguistic features, we evaluate the output quality of CrossSpeech++ trained with linguistic features from different layers of MMS~\cite{pratap2024scaling}. 
Specifically, we compare the linguistic features extracted from the 1$^\text{st}$, 6$^\text{th}$, 12$^\text{th}$, 18$^\text{th}$, and 24$^\text{th}$ layers.
Table~\ref{tab:SSL} presents the evaluation results on our validation sets, indicating trade-offs across different layers. 
When we inject hidden features from the earlier layers into LDG, it brings about language-speaker entanglement, resulting in the text embeddings learning pronunciation along with the corresponding native speaker information.
While this leads to more intelligible speech (measured by CER), it results in degraded naturalness (measured by UTMOS) and speaker similarity (measured by SECS), which is not a desired outcome.
However, when we utilize the hidden features from the latter layers, it contributes more to language-speaker disentanglement, leading to improved naturalness and speaker similarity.
Therefore, we use the features from the 24$^\text{th}$ layer because it provides improved naturalness and speaker similarity with a slight reduction in intelligibility.

\begin{table}[t]
\centering
\caption{Evaluation on different configurations of linguistic feature extraction. \# represents the layer index of MMS~\cite{pratap2024scaling}.}
\resizebox{0.99\columnwidth}{!}
{
\begin{tabular}{lcccccc}
\toprule
\multirow{2}{*}{\vspace{-1mm}\bfseries ~\#} & \multicolumn{3}{c}{\bfseries Cross-lingual} & \multicolumn{3}{c}{\bfseries Intra-lingual} \\ \cmidrule(lr){2-4}\cmidrule(lr){5-7}

    &UTMOS$\uparrow$  &SECS$\uparrow$ &CER$\downarrow$ &UTMOS$\uparrow$ &SECS$\uparrow$ &CER$\downarrow$    \\  \cmidrule(lr){1-7}
    ~1   &$3.672$ &$0.710$ &$\bf{12.45}$ &$3.274$ &$0.774$ &~$\bf{9.42}$    \\
    ~6   &$3.796$ &$0.743$ &$12.99$ &$3.343$  &$0.774$ &$10.13$  \\
    12  &$3.799$ &$0.748$ &$13.10$ &$\bf{3.369}$  &$0.776$ &~$9.68$  \\
    18  &$3.800$ &$0.747$ &$13.01$ &$3.365$   &$0.777$ &~$9.77$ \\
    24  &$\bf{3.829}$ &$\bf{0.756}$ &$13.58$ &$\bf{3.369}$    &$\bf{0.778}$ &$10.10$ \\
\bottomrule
\end{tabular}
}
\label{tab:SSL}
\end{table}

\subsection{Ablation Study}
We investigate the effect of each CrossSpeech++ component by conducting an ablation study on its quality.
We measure UTMOS, SECS, and CER in both cross-lingual and intra-lingual cases.
As indicated in Table~\ref{tab:ablation}, each component contributes to enhancing the quality of CrossSpeech++. 
{Replacing MDSLN with the original LN in the LD encoder (\textit{w/o} MDSLN) results in relatively small yet consistent degradation across all metrics in both cross-lingual and intra-lingual cases. 
This indicates that MDSLN helps to learn speaker-generalizable features and facilitates the training of the subsequent LDV adaptor.}
Moreover, the Linguistic Adaptor (LA) significantly improves naturalness and intelligibility in both cross-lingual and intra-lingual cases.
While it slightly reduces speaker similarity, we presume this is due to residual speaker information entangled within the text representations.
Removing audio perturbation hinders the effectiveness of LA in disentangling language and speaker information, resulting in noticeable degradation across all metrics.
The absence of the Text Predictor (TP) when extracting target linguistic features also leads to inaccurate pronunciation.
The importance of modeling LD and SD speech variations {(\textit{w/o} LDV and \textit{w/o} SDV )} is validated by the degraded quality observed when these variations are overlooked.

\begin{table}[t]
\centering
\caption{Results for the ablation study. LA and TP refer to linguistic adaptor and text predictor, respectively.}
\resizebox{0.99\columnwidth}{!}
{
\begin{tabular}{lcccccc}
\toprule
\multirow{2}{*}{\vspace{-1mm}\bfseries Method} & \multicolumn{3}{c}{\bfseries Cross-lingual} & \multicolumn{3}{c}{\bfseries Intra-lingual} \\ \cmidrule(lr){2-4}\cmidrule(lr){5-7}
    &UTMOS$\uparrow$  &SECS$\uparrow$ &CER$\downarrow$ &UTMOS$\uparrow$ &SECS$\uparrow$ &CER$\downarrow$ \\ \cmidrule(lr){1-7}
    CrossSpeech++ &$\bf 3.791$ &$0.761$ &$\bf 13.35$ &$\bf 3.434$ &$0.777$ &$\bf 10.14$     \\ \cmidrule{1-7}
    \emph{w/o} MDSLN  &$3.767$ &$0.752$ &$13.39$ &$3.422$  &$0.762$ &$11.32$  \\
    \emph{w/o} LDV    &$3.763$ &$0.750$ &$13.43$ &$3.346$  &$0.770$  &$\bf 10.14$ \\
    \emph{w/o} LA  &$3.443$ &$\bf 0.772$ &$13.54$ &$3.115$  &$\bf 0.783$ &$10.53$  \\
    \emph{w/o} Perturb  &$3.611$ &$0.706$ &$15.58$ &$3.343$  &$0.768$ &$10.43$  \\
    \emph{w/o} TP  &$3.782$ &$0.751$ &$13.42$ &$3.311$  &$0.772$  &$\bf 10.13$ \\
    \emph{w/o} SDV  &$3.695$ &$0.753$ &$14.02$ &$3.227$ &$0.765$ &$10.93$ \\
\bottomrule
\end{tabular}
}
\label{tab:ablation}
\end{table}

\begin{figure*}[ht]
    \centering
    \includegraphics[width=0.9\textwidth]{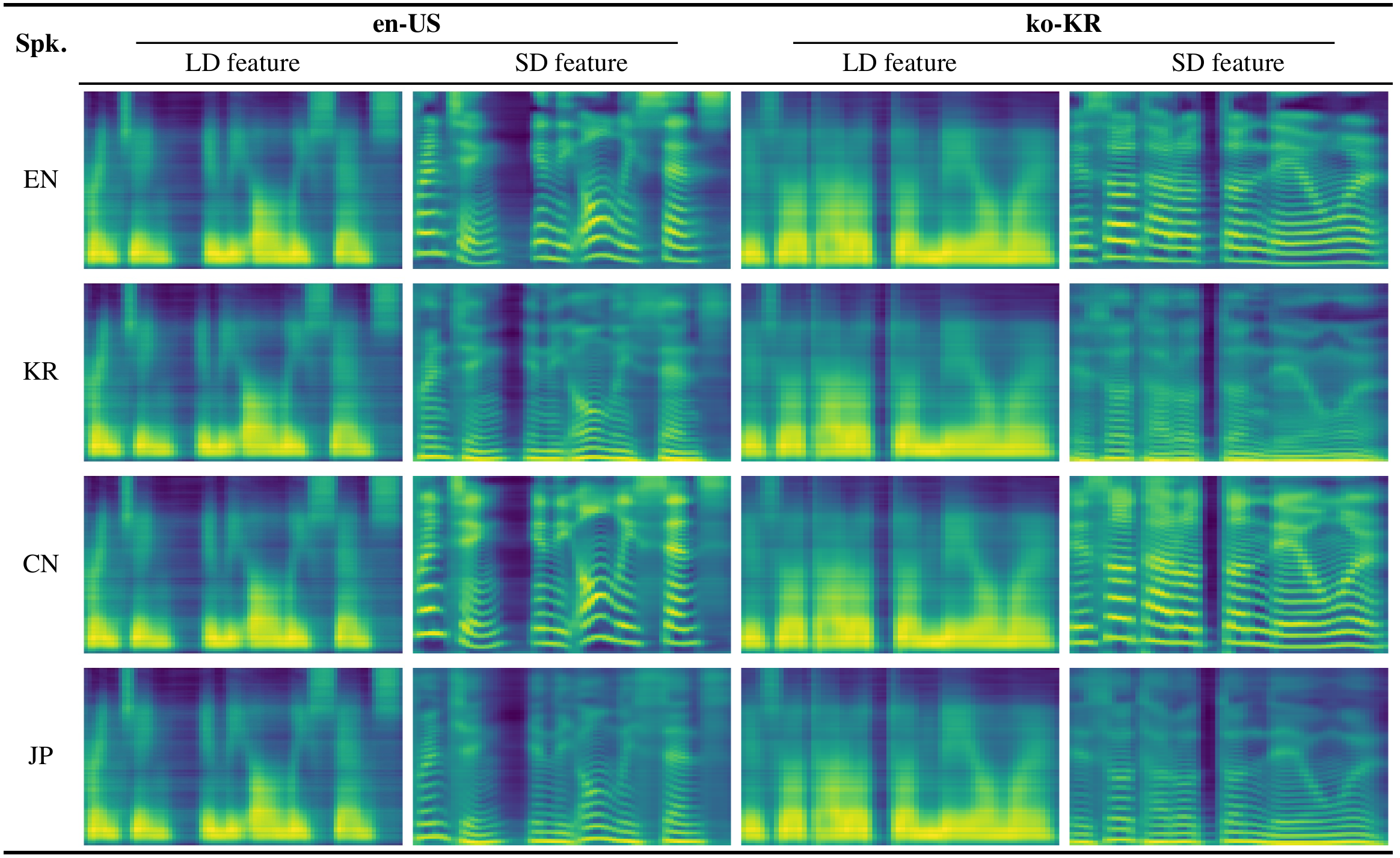}
    \caption{Visualization of language-dependent (LD) and speaker-dependent (SD) features.
    We visualize LD and SD features based on two different languages (en-Us and ko-KR) and spoken by four different speakers, i.e., English (EN), Korean (KR), Chinese (CN), and Japaneses (JP). 
    Note that the LD feature remains invariant, while the SD feature varies across different speakers.
    }
    \vspace{-5mm}
    \label{fig:feat_mel}
\end{figure*}

\begin{figure}[t]
    \centering
    \includegraphics[width=0.9\columnwidth]{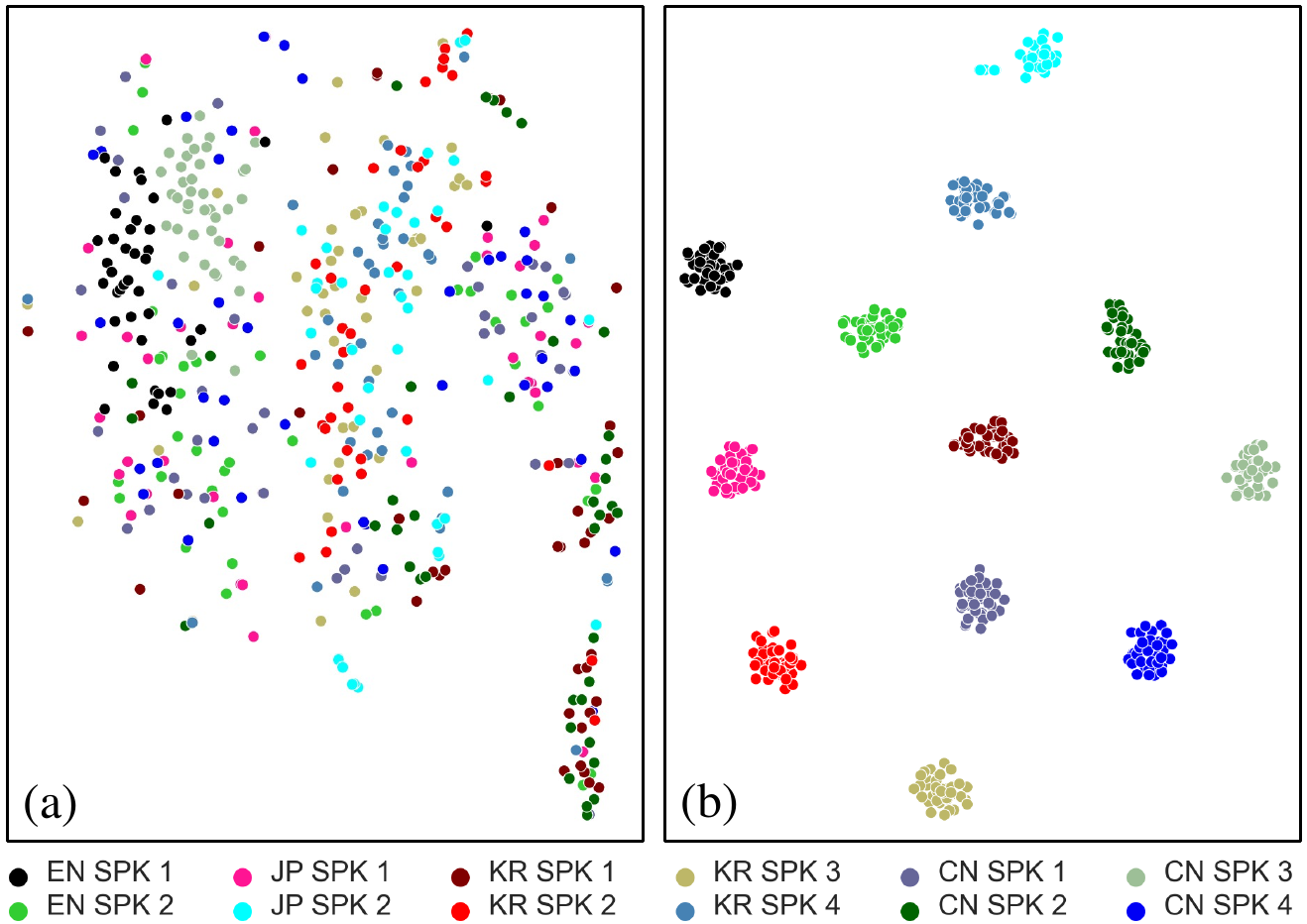}
    \caption{t-SNE plots of speaker feature space of (a) LD features and (b) the output mel-spectrogram. 
    Each color represents different speakers.
    }
    \vspace{-3.5mm}
    \label{fig:feat_space}
\end{figure}

\subsection{Qualitative Evaluation}
To intuitively demonstrate speaker generalization capability of LDG and speaker transferability of SDG, we visualize the LD and SD features.
Fig.~\ref{fig:feat_mel} illustrates the LD and SD features derived from text inputs in two different languages (en-US and ko-KR) and spoken by four different speakers (EN, KR, CN, and JP).
As evident from the figure, the LD feature does not contain speaker-specific characteristics (e.g., harmonics) and changes only according to the input text regardless of speaker information.
On the contrary, the SD feature includes speaker-specific characteristics and varies with different speakers.
This indicates that CrossSpeech++ successfully disentangles language and speaker-related information into acoustic representations, each dependent on the corresponding information.

In addition, Fig.~\ref{fig:feat_space} depicts the speaker feature space of (a) LD features and (b) the out mel-spectrogram by using t-Stochastic Neighbor embedding (t-SNE)~\cite{van2008visualizing}.
In Fig.~\ref{fig:feat_space}(a), we observe that the embeddings are not clustered by speakers but rather randomly spread out. 
This indicates that the language-dependent representations are not biased to speaker-related information but solely contain text-related variations.
On the other hand, the embeddings are well-clustered by {speakers} in Fig.~\ref{fig:feat_space}(b), demonstrating CrossSpeech++ can successfully learn and transfer speaker-dependent characteristics through SDG.

\begin{table}[t]
\centering
\caption{{Quality comparison with zero-shot cross-lingual models.}}
\vspace{-2mm}
{
\begin{tabular}{lcccc}
\toprule
    \textbf{Method}  &UTMOS$\uparrow$ &SECS$\uparrow$ &CER$\downarrow$ & \\  \cmidrule(lr){1-5}
    VALL-E X~\cite{zhang2023speak} &3.243 &0.710 &34.76 & \\
    XTTS-v2~\cite{casanova2024xtts} &3.450 &{0.763} &\textbf{17.80} & \\  \cmidrule(lr){1-5}
    \textbf{CrossSpeech++}  &\textbf{3.863} &\textbf{0.767} &{21.22} & \\
\bottomrule
\end{tabular}
}
\vspace{-5mm}
\label{tab:zero}
\end{table}

\subsection{{Comparison with Zero-shot Models}}
{
We further evaluate our method in comparison to recent zero-shot cross-lingual models: VALL-E X~\cite{zhang2023speak} and XTTS-v2~\cite{casanova2024xtts}. 
Using the pre-trained checkpoints from the popular reproduction of VALL-E X\footnote{\url{https://github.com/Plachtaa/VALL-E-X}} and the official implementation of XTTS-v2\footnote{\url{https://github.com/coqui-ai/TTS}}, we generate audio samples in a zero-shot manner and compute UTMOS, SECS, and CER. 
Different from the experiments in Table~\ref{tab:compare} where we evaluate using all languages, we focus on English, Chinese, and Japanese sentences in this evaluation, as VALL-E X does not support Korean synthesis. 
The evaluation results are presented in Table~\ref{tab:zero}. 
Although CrossSpeech++ exhibits a slightly higher CER than XTTS-v2, our method consistently outperforms all zero-shot baselines in terms of UTMOS and SECS. This demonstrates that our approach generates more natural-sounding speech with accurate speaker characteristics.
}

\section{Broader Impact}
By leveraging CrossSpeech++, we can achieve various positive societal impacts, such as creating educational resources for foreign language learning and developing conversational AI agents with multilingual capabilities, all while preserving a consistent speaker identity. 
However, it is crucial to recognize the potential threats that could arise from the misuse of this technology.
These threats include the creation of hate speech and voice phishing attacks. 
Additionally, the ability to convert text to speech in multiple languages poses a risk of spreading misinformation globally in one's own voice, thus amplifying its reach and impact. 
These considerations highlight the necessity of responsible use and the establishment of ethical guidelines in the deployment of cross-lingual TTS systems.

\section{Conclusion and Discussion}
In this paper, we propose CrossSpeech++, which achieves high-fidelity cross-lingual speech synthesis with significantly improved speech naturalness.
We observed remain language-speaker disentanglement in previous cross-lingual TTS systems and addressed the issue by separately modeling language and speaker representations in the output acoustic features.
Experimental results demonstrated that CrossSpeech++ outperformed standard methods both in cross-lingual and intra-lingual scenarios.
Moreover, we verified the effectiveness of each CrossSpeech component by conducting an ablation study. 

CrossSpeech++ has demonstrated remarkable capabilities in synthesizing both cross- and intra-lingual speech compared to previous works. 
However, despite its advancements, CrossSpeech++ requires a substantial corpus of text-to-speech pairs to produce speech in a target language, making it less applicable to low-resource languages. 
Therefore, our future research will focus on developing effective strategies to deploy cross-lingual TTS systems, even in low-resource language.

\bibliographystyle{IEEEtran}
\bibliography{main}

\vspace{-15mm}
\begin{IEEEbiographynophoto}{Ji-Hoon Kim} is a Ph.D. student in Electrical Engineering at the Korea Advanced Institute of Science and Technology, Daejeon, Republic of Korea.
His main research interests include speech processing and multi-modal learning.
He received the M.S. degree in Artificial Intelligence from the Korea University, Seoul, Republic of Korea.
\end{IEEEbiographynophoto}
\vspace{-10mm}

\vspace{-8mm}
\begin{IEEEbiographynophoto}{Hong-Sun Yang}
is an AI Engineer at 42dot, Seoul, Republic of Korea.
His main research interests include speech synthesis.
He received the M.S. degree from Korea University, Seoul, Republic of Korea.
\end{IEEEbiographynophoto}
\vspace{-10mm}

\vspace{-8mm}
\begin{IEEEbiographynophoto}{Yoon-Cheol Ju}
is an AI Engineer at 42dot, Seoul, Republic of Korea. 
His main research interests include speech synthesis.
He received the bachelor's degree from Sogang University, Seoul, Republic of Korea. 
\end{IEEEbiographynophoto}
\vspace{-10mm}

\vspace{-8mm}
\begin{IEEEbiographynophoto}{Il-Hwan Kim}
is an AI Engineer at 42dot, Seoul, Republic of Korea.
His main research interests include zero-shot speech synthesis, the generation of sound effects, and applications of speech synthesis.
He received the M.S. in Electronic Engineering from Kyungpook National University. 
\end{IEEEbiographynophoto}
\vspace{-10mm}

\vspace{-8mm}
\begin{IEEEbiographynophoto}{Byeong-Yeol Kim} 
is the group lead of audio group at 42dot.
His main research interests include speech recognition, speech synthesis, and speech signal processing.
He received the M.S. in Electrical Engineering from Korea Advanced Institute of Science and Technology, Daejeon, South Korea.
\end{IEEEbiographynophoto}
\vspace{-10mm}

\vspace{-8mm}
\begin{IEEEbiographynophoto}{Joon Son Chung}
is an assistant professor at Korea Advanced Institute of Science and Technology, where he is directing research in speech processing, computer vision and machine learning. 
He received the D.Phil. in Engineering Science from the University of Oxford.
\end{IEEEbiographynophoto}

\end{document}